\documentclass[epjCONF]{svjour}

\usepackage{graphics}
\usepackage[varg]{txfonts} 
\usepackage[latin1]{inputenc}

\RequirePackage{longtable}
\RequirePackage{tabularx}
\RequirePackage{dcolumn}

\newcommand{\mev}{\mbox{MeV}}

\newcommand{\be}{\begin{equation}}
\newcommand{\ee}{\end{equation}}
\newcommand{\bea}{\begin{eqnarray}}
\newcommand{\eea}{\end{eqnarray}}
\newcommand{\bdm}{\begin{displaymath}}
\newcommand{\edm}{\end{displaymath}}

\session-title{MESON2012 - 12th International Workshop on Meson
Production, Properties and Interaction}

\begin{document}
\title{Hadronic light-by-light scattering in the muon $g-2$: impact of
   proposed measurements of the $\pi^0\to\gamma\gamma$ decay width and
   the $\gamma^\ast\gamma\to\pi^0$ transition form factor with the
   KLOE-2 experiment}
\author{Andreas
  Nyffeler\thanks{\email{nyffeler@hri.res.in}}}
\institute{Regional Centre for Accelerator-based Particle Physics,
  Harish-Chandra Research Institute \\
  Chhatnag Road, Jhusi, Allahabad - 211019, India} 
\abstract{
We discuss, how planned measurements at KLOE-2 of the
$\pi^0\to\gamma\gamma$ decay width and the $\gamma^\ast\gamma\to\pi^0$
transition form factor can improve estimates for the numerically
dominant pion-exchange contribution to hadronic light-by-light
scattering in the muon $g-2$ and what are the limitations related to
the modelling of the off-shellness of the pion.
} 
\maketitle


\section{Introduction}
\label{intro}

The anomalous magnetic moment of the muon $a_\mu$ provides an
important test of the Standard Model (SM) and is potentially sensitive
to contributions from New Physics~\cite{JN_09}. In fact, for several
years now a deviation is observed between the experimental measurement
and the SM prediction, $a_\mu^{\rm exp} - a_\mu^{\rm SM} \sim
(250-300) \times 10^{-11}$, corresponding to about $3 - 3.5$~standard
deviations~\cite{JN_09,recent_estimates}.

Hadronic effects dominate the uncertainty in the SM prediction of
$a_\mu$ and make it difficult to interpret this discrepancy as a sign
of New Physics. In particular, in contrast to the hadronic vacuum
polarization in the $g-2$, which can be related to data, the estimates
for the hadronic light-by-light (LbyL) scattering contribution
$a_\mu^{{\rm had.\ LbyL}} = (105 \pm 26) \times
10^{-11}$~\cite{PdeRV_09} and $a_\mu^{{\rm had.\ LbyL}} = (116 \pm 40)
\times 10^{-11}$~\cite{Nyffeler_09,JN_09} rely entirely on
calculations using {\it hadronic models} which employ form factors for
the interaction of hadrons with photons. The more recent
papers~\cite{Schwinger_Dyson,GdeR_12} yield a larger central value and
a larger error of about $(150 \pm 50) \times 10^{-11}$. For a brief
review of had.\ LbyL scattering in $a_\mu$ see Ref.~\cite{B_ZA}, which
also includes a reanalysis of the charged pion loop contribution, see
also Ref.~\cite{pion_loop}.  To fully profit from future planned $g-2$
experiments with a precision of $15 \times 10^{-11}$, these large
model uncertainties have to be reduced. Maybe lattice QCD will at some
point give a reliable number, see the talk~\cite{Blum_Lattice_2012}
for some encouraging progress recently. Meanwhile, experimental
measurements and theoretical constraints of the relevant form factors
can help to constrain the models and to reduce the uncertainties in
$a_\mu^{{\rm had.\ LbyL}}$.

In most model calculations, pion-exchange gives the numerically
dominant contribution\footnote{Apart from the recent
  papers~\cite{Schwinger_Dyson,GdeR_12}, where the (dressed) quark
  loop gives the largest contribution.}, therefore it has received a
lot of attention. In our paper~\cite{KLOE-2_impact} we studied the
impact of planned measurements at KLOE-2 of the $\pi^0\to\gamma\gamma$
decay width to 1\% statistical precision and the
$\gamma^\ast\gamma\to\pi^0$ transition form factor ${\cal
  F}_{\pi^0\gamma^\ast\gamma}(Q^2)$ for small space-like momenta,
$0.01~\mbox{GeV}^2 \leq Q^2 \leq 0.1~\mbox{GeV}^2$, to 6\% statistical
precision in each bin, on estimates of the pion-exchange contribution
$a_\mu^{{\rm LbyL}; \pi^0}$.  We would like to stress that a realistic
calculation of $a_\mu^{{\rm LbyL}; \pi^0}$ is {\it not} the purpose of
this paper.  The estimates given below are performed to demonstrate,
within several models, an improvement of uncertainty, which will be
possible when the KLOE-2 data appear. The simulations in
Ref.~\cite{KLOE-2_impact} have been performed with the dedicated
Monte-Carlo program EKHARA~\cite{EKHARA} for the process $e^+ e^- \to
e^+ e^- \gamma^* \gamma^* \to e^+ e^- P$ with $P = \pi^0, \eta,
\eta^\prime$, followed by the decay $\pi^0 \to \gamma\gamma$ and
combined with a detailed detector simulation.

\section{Impact of KLOE-2 measurements on $a_\mu^{\mathrm{LbyL};\pi^0}$}
\label{sec:KLOE}

Any experimental information on the neutral pion lifetime and the
transition form factor is important in order to constrain the models
used for calculating the pion-exchange contribution.  However, having
a good description e.g.\ for the transition form factor is only
necessary, not sufficient, in order to uniquely determine
$a_\mu^{\mathrm{LbyL};\pi^0}$. As stressed in
Ref.~\cite{Jegerlehner_off-shell}, what enters in the calculation of
$a_\mu^{{\rm LbyL}; \pi^0}$ is the fully off-shell form factor ${\cal
  F}_{{\pi^0}^*\gamma^*\gamma^*}((q_1 + q_2)^2, q_1^2, q_2^2)$ (vertex
function), where also the pion is off-shell with 4-momentum $(q_1 +
q_2)$. Such a (model dependent) form factor can for instance be
defined via the QCD Green's function $\langle VVP \rangle$, see
Ref.~\cite{Nyffeler_09} for details and references to earlier work.
The form factor with on-shell pions is then given by ${\cal
  F}_{\pi^0\gamma^*\gamma^*}(q_1^2, q_2^2) \equiv {\cal
  F}_{{\pi^0}^*\gamma^*\gamma^*}(m_\pi^2, q_1^2, q_2^2)$.
Measurements of the transition form factor ${\cal
  F}_{\pi^0\gamma^\ast\gamma}(Q^2) \equiv {\cal
  F}_{{\pi^0}^\ast\gamma^\ast\gamma^\ast}(m_{\pi}^2, -Q^2, 0)$ are in
general only sensitive to a subset of the model parameters and do not
allow to reconstruct the full off-shell form factor.

For different models, the effects of the off-shell pion can vary a
lot. In Ref.~\cite{Nyffeler_09} an off-shell form factor (LMD+V) was
proposed, based on large-$N_C$ QCD matched to short-distance
constraints from the operator product expansion, see also
Ref.~\cite{KN_EPJC_2001}. This yields the estimate $a_{\mu; {\rm
    LMD+V}}^{{\rm LbyL}; \pi^0} = (72 \pm 12) \times 10^{-11}$. The
error estimate comes from the variation of all model parameters, where
the uncertainty of the parameters related to the off-shellness of the
pion completely dominates the total error and will {\it not} be shown
in Table~\ref{tab:amu} below.

In contrast to the off-shell LMD+V model, many models, e.g.\ the VMD
model, constituent quark models or the ans\"atze for the transition
form factor used in Ref.~\cite{Cappiello:2010uy}, do not have these
additional sources of uncertainty related to the off-shellness of the
pion. These models often have only very few parameters, which can all
be fixed by measurements of the transition form factor or from other
observables. Therefore, the precision of the KLOE-2 measurement can
dominate the total accuracy of $a_\mu^{\mathrm{LbyL};\pi^0}$ in such
models.

It was noted in Ref.~\cite{Nyffeler:2009uw} that essentially all
evaluations of the pion-exchange contribution use the normalization
${\cal F}_{{\pi^0}^*\gamma^*\gamma^*}(m_\pi^2, 0, 0) = 1 / (4 \pi^2
F_\pi)$ for the form factor, as derived from the Wess-Zumino-Witten
(WZW) term. Then the value $F_\pi = 92.4~\mev$ is used without any
error attached to it, i.e. a value close to $F_\pi = (92.2 \pm
0.14)~\mbox{MeV}$, obtained from $\pi^+ \to \mu^+
\nu_\mu(\gamma)$~\cite{Nakamura:2010zzi}. Instead, if one uses the
decay width $\Gamma_{\pi^0 \to \gamma\gamma}$ for the normalization of
the form factor, an additional source of uncertainty enters, which has
not been taken into account in most evaluations.

In our calculations we account for this normalization issue, using in 
the fit: 
\begin{itemize}
\item $\Gamma^{{\rm PDG}}_{\pi^0 \to \gamma\gamma} = 7.74 \pm
0.48$~eV from the PDG 2010~\cite{Nakamura:2010zzi},
\item $\Gamma^{{\rm PrimEx}}_{\pi^0 \to \gamma\gamma} = 7.82 \pm
0.22$~eV from the PrimEx experiment~\cite{Larin:2010kq},
\item $\Gamma^{{\rm KLOE-2}}_{\pi^0 \to \gamma\gamma} = 7.73 \pm
0.08$~eV for the KLOE-2 simulation (assuming a $1\%$ precision). 
\end{itemize}

The assumption that the KLOE-2 measurement will be consistent with the
LMD+V and VMD models, allowed us in Ref.~\cite{KLOE-2_impact} to use
the simulations as new ``data'' and evaluate the impact of such
``data'' on the precision of the $a_\mu^{{\rm LbyL}; \pi^0}$
calculation.  In order to do that, we fit the LMD+V and VMD models to
the data sets~\cite{TFF_data} from CELLO, CLEO and BaBar for the
transition form factor and the values for the decay width given above:
\bdm 
\label{eq:fitdatasets}
 \begin{array}{ll}
   A0: & \mbox{CELLO, CLEO, PDG} \\
   A1: & \mbox{CELLO, CLEO, PrimEx} \\
   A2: & \mbox{CELLO, CLEO, PrimEx, KLOE-2} \\
   B0: & \mbox{CELLO, CLEO, BaBar, PDG} \\
   B1: & \mbox{CELLO, CLEO, BaBar, PrimEx} \\
   B2: & \mbox{CELLO, CLEO, BaBar, PrimEx, KLOE-2}
 \end{array}
\edm

The BaBar measurement of the transition form factor does not show the
$1/Q^2$ behavior as expected from earlier theoretical considerations
by Brodsky-Lepage~\cite{Brodsky-Lepage} and as seen in the CELLO and
CLEO data and the recent measurements from Belle~\cite{Belle}.  The
VMD model always shows a $1/Q^2$ fall-off and therefore is not
compatible with the BaBar data. The LMD+V model has another parameter,
$h_1$, which determines the behavior of the transition form factor for
large $Q^2$. To get the $1/Q^2$ behavior according to Brodsky-Lepage,
one needs to set $h_1 = 0$. However, one can simply leave $h_1$ as a
free parameter and fit it to the BaBar data, yielding $h_1 \neq
0$~\cite{Nyffeler:2009uw}. In this case the form factor does not
vanish for $Q^2 \to \infty$. Since VMD and LMD+V with $h_1 = 0$ are
not compatible with the BaBar data, the corresponding fits are very
bad and we will not include these results in the current paper, see
Ref.~\cite{KLOE-2_impact} for details.

For illustration, we use the following two approaches to calculate
 $a_\mu^{{\rm LbyL}; \pi^0}$: 
\begin{itemize}
 \item Jegerlehner-Nyffeler (JN)
       approach~\cite{Nyffeler_09,JN_09} 
       with the off-shell pion form factor;
 \item Melnikov-Vainshtein (MV) approach~\cite{Melnikov:2003xd},
       where one uses the on-shell pion form factor at the internal
       vertex and a constant (WZW) form factor at the external vertex.
\end{itemize}

Table~\ref{tab:amu} shows the impact of the PrimEx and the future
KLOE-2 measurements on the model parameters and, consequently, on the
$a_\mu^{{\rm LbyL}; \pi^0}$ uncertainty. The other parameters of the
(on-shell and off-shell) LMD+V model have been chosen as in the
papers~\cite{Nyffeler_09,JN_09,Melnikov:2003xd}.  We stress again that
our estimate of the $a_\mu^{{\rm LbyL}; \pi^0}$ uncertainty is given
only by the propagation of the errors of the fitted parameters in
Table~\ref{tab:amu} and therefore we may not reproduce the total
uncertainties given in the original papers.
 
\begin{table*}
 \caption{KLOE-2 impact on the accuracy of $a_\mu^{{\rm
       LbyL}; \pi^0}$ in case of one year of data taking
   ($5$~fb$^{-1}$). The values marked with asterisk (*)
   do not contain additional uncertainties coming from
   the ``off-shellness'' of the pion (see the text).} 
 \label{tab:amu}
 {\scriptsize 
 \begin{tabularx}{\textwidth}{clcllll}
 Model&Data& $\chi^2/d.o.f.$ &  & Parameters && $a_\mu^{{\rm LbyL};
   \pi^0} \times 10^{11}$\\ 
 \hline
 VMD  & A0 & $6.6/19$
     & $M_V = 0.778(18)$~GeV & $F_\pi = 0.0924(28)$~GeV  && $(57.2 \pm
 4.0)_{JN}$\\ 
 VMD  & A1 & $6.6/19$
     & $M_V = 0.776(13)$~GeV & $F_\pi = 0.0919(13)$~GeV  && $(57.7 \pm
 2.1)_{JN}$\\ 
 VMD  & A2 & $7.5/27$
     & $M_V = 0.778(11)$~GeV & $F_\pi = 0.0923(4)$~GeV   && $(57.3 \pm
 1.1)_{JN}$\\ 
 \hline
 LMD+V, $h_1 = 0$  & A0 & $6.5/19$ 
     &  $\bar{h}_5 = 6.99(32)$~GeV$^4$ & $\bar{h}_7 =
 -14.81(45)$~GeV$^6$ && $(72.3 \pm 3.5)_{JN}^*$\\ 
 &  &   &                        &                            &&
 $(79.8 \pm 4.2)_{MV}$\\ 
 LMD+V, $h_1 = 0$  & A1 & $6.6/19$ 
     &  $\bar{h}_5 = 6.96(29)$~GeV$^4$ & $\bar{h}_7 =
 -14.90(21)$~GeV$^6$ && $(73.0 \pm 1.7)_{JN}^*$\\ 
 &  &   &                        &                            &&
 $(80.5 \pm 2.0)_{MV}$\\ 
 LMD+V, $h_1 = 0$  & A2 & $7.5/27$
     &  $\bar{h}_5 = 6.99(28)$~GeV$^4$ & $\bar{h}_7 =
 -14.83(7)$~GeV$^6$ && $(72.5 \pm 0.8)_{JN}^*$\\ 
 &  &   &                        &                           && $(80.0
 \pm 0.8)_{MV}$\\ 
 \hline
 LMD+V, $h_1 \neq 0$  & A0 & $6.5/18$ 
     &  $\bar{h}_5 = 6.90(71)$~GeV$^4$ & $\bar{h}_7 = -14.83(46)$~GeV$^6$& 
        $h_1 = -0.03(18)$~GeV$^2$ & $(72.4 \pm 3.8)_{JN}^*$\\
 LMD+V, $h_1 \neq 0$  & A1 & $6.5/18$ 
     &  $\bar{h}_5 = 6.85(67)$~GeV$^4$ & $\bar{h}_7 = -14.91(21)$~GeV$^6$& 
        $h_1 = -0.03(17)$~GeV$^2$ & $(72.9 \pm 2.1)_{JN}^*$\\
 LMD+V, $h_1 \neq 0$  & A2 & $7.5/26$ 
     &  $\bar{h}_5 = 6.90(64)$~GeV$^4$ & $\bar{h}_7 = -14.84(7)$~GeV$^6$ &
        $h_1 = -0.02(17)$~GeV$^2$ & $(72.4 \pm 1.5)_{JN}^*$\\
 LMD+V, $h_1 \neq 0$  & B0 & $18/35$ 
     &  $\bar{h}_5 = 6.46(24)$~GeV$^4$ & $\bar{h}_7 = -14.86(44)$~GeV$^6$ &
        $h_1 = -0.17(2)$~GeV$^2$ & $(71.9 \pm 3.4)_{JN}^*$\\
 LMD+V, $h_1 \neq 0$  & B1 & $18/35$ 
     &  $\bar{h}_5 = 6.44(22)$~GeV$^4$ & $\bar{h}_7 = -14.92(21)$~GeV$^6$ &
        $h_1 = -0.17(2)$~GeV$^2$ & $(72.4 \pm 1.6)_{JN}^*$\\
 LMD+V, $h_1 \neq 0$  & B2 & $19/43$ 
     &  $\bar{h}_5 = 6.47(21)$~GeV$^4$ & $\bar{h}_7 = -14.84(7)$~GeV$^6$ &
        $h_1 = -0.17(2)$~GeV$^2$ & $(71.8 \pm 0.7)_{JN}^*$ \\
\hline
 \end{tabularx}
}
\end{table*}

We can clearly see from Table~\ref{tab:amu} that for each given model
and each approach (JN or MV), there is a trend of reduction in the
error for $a_\mu^{{\rm LbyL}; \pi^0}$ (related only to the given model
parameters) by about half when going from A0 (PDG) to A1 (including
PrimEx) and by about another half when going from A1 to A2 (including
KLOE-2). Very roughly, we can write:
\begin{itemize}
\item 
Sets A0, B0: $\delta a_\mu^{{\rm LbyL}; \pi^0} \approx 4 \times
10^{-11}$ (with $\Gamma^{{\rm PDG}}_{\pi^0 \to \gamma\gamma}$)
\item 
Sets A1, B1: $\delta a_\mu^{{\rm LbyL}; \pi^0} \approx2 \times
10^{-11}$
(with $\Gamma^{{\rm PrimEx}}_{\pi^0 \to \gamma\gamma}$) 
\item   
Sets A2, B2: $\delta a_\mu^{{\rm LbyL}; \pi^0} \approx (0.7 - 1.1)
\times 10^{-11}$ (with simulated KLOE-2 data)
\end{itemize}

This is mainly due to the improvement in the normalization of the form
factor, related to the decay width $\pi^0 \to \gamma\gamma$,
controlled by the parameters $F_\pi$ or $\bar{h}_7$, respectively, but
more data also better constrain the other model parameters $M_V$ or
$\bar{h}_5$. This trend of improvement is also visible in the last
part of the Table (LMD+V, $h_1 \neq 0$), when we fit the sets B0, B1
and B2 which include the BaBar data.  The central values of the final
results for $a_\mu^{{\rm LbyL}; \pi^0}$ are only slightly changed, if
we include the BaBar data.  They shift only by about $-0.5 \times
10^{-11}$ compared to the corresponding data sets A0, A1 and A2.  This
is due to a partial compensation in $a_\mu^{{\rm LbyL}; \pi^0}$, when
the central values for $\bar{h}_5$ and $h_1$ are changed, see
Ref.~\cite{Nyffeler:2009uw}.

Finally, note that both VMD and LMD+V with $h_1 = 0$ can fit the data
sets A0, A1 and A2 for the transition form factor very well with
essentially the same $\chi^2$ per degree of freedom for a given data
set (see first and second part of the table).  Nevertheless, the
results for the pion-exchange contribution differ by about $20\%$ in
these two models. For VMD the result is $a_\mu^{{\rm LbyL}; \pi^0}
\sim 57.5 \times 10^{-11}$ and for LMD+V with $h_1 = 0$ it is $72.5
\times 10^{-11}$ with the JN approach and $80 \times 10^{-11}$ with
the MV approach. This is due to the different behavior, in these two
models, of the fully off-shell form factor ${\cal
  F}_{{\pi^0}^*\gamma^*\gamma^*}((q_1 + q_2)^2, q_1^2, q_2^2)$ on all
momentum variables, which enters for the pion-exchange
contribution~\cite{Jegerlehner_off-shell}. The VMD model is known to
have a wrong high-energy behavior with too strong damping, which
underestimates the contribution. For the VMD model, measurements of
the neutral pion decay width and the transition form factor completely
determine the model parameters $F_\pi$ and $M_V$ and the error given
in Table~\ref{tab:amu} is the total model error. Note that a smaller
error, compared to the off-shell LMD+V model, does not necessarily
imply that the VMD model is better, i.e.\ closer to reality. Maybe the
model is too simplistic.

We conclude that the KLOE-2 data with a total integrated luminosity of
$5$~fb$^{-1}$ will give a reasonable improvement in the part of the
$a_\mu^{{\rm LbyL}; \pi^0}$ error associated with the parameters
accessible via the $\pi^0 \to \gamma\gamma$ decay width and the
$\gamma^\ast\gamma \to \pi^0$ transition form factor. Depending on the
modelling of the off-shellness of the pion, there might be other,
potentially larger sources of uncertainty which cannot be improved by
the KLOE-2 measurements.

\section*{Acknowledgements}

I would like to thank my coauthors on Ref.~\cite{KLOE-2_impact} for
the pleasant collaboration on this work. I am grateful to the
organizers of MESON 2012 for the opportunity to present our paper and
for providing a stimulating atmosphere during the meeting. This work
is supported by funding from the Department of Atomic Energy,
Government of India, for the Regional Centre for Accelerator-based
Particle Physics (RECAPP), Harish-Chandra Research Institute.

\end{document}